\documentclass[10pt,conference]{IEEEtran}
\IEEEoverridecommandlockouts
\usepackage{cite}
\usepackage{amsmath,amssymb,amsfonts}
\usepackage{algorithmic}
\usepackage{booktabs}
\usepackage{makecell}  % 表格内换行
\usepackage{array}     % 表格列定义
\usepackage{tabularx}  % 自动换行
\usepackage{graphicx}
\usepackage{textcomp}
\usepackage{xcolor}
\def\BibTeX{{\rm B\kern-.05em{\sc i\kern-.025em b}\kern-.08em
    T\kern-.1667em\lower.7ex\hbox{E}\kern-.125emX}}
\newcommand{\ourtool}{$CQ^2A$}
\newcommand{\qaqa}{QAQA}
\newcommand{\llm}{LLM}
\newcommand{\chatgpt}{ChatGPT}
\newcommand{\simcse}{SimCSE}
\newcommand{\squad}{SQuAD2}
\usepackage{subcaption}
\usepackage{multirow}
\usepackage{hyperref}
\hypersetup{hidelinks,
 colorlinks=true,
 allcolors=black,
 pdfstartview=Fit,
 breaklinks=true}
\begin{document}

\title{Testing Question Answering Software with Context-Driven Question Generation}

\author{
\IEEEauthorblockN{Shuang Liu}
\IEEEauthorblockA{\textit{Renmin University of China} \\
shuang.liu@ruc.edu.cn}
\and
\IEEEauthorblockN{Zhirun Zhang}
\IEEEauthorblockA{\textit{Tianjin University} \\
zhangzhirun2000@sina.com}
\and
\IEEEauthorblockN{Jinhao Dong}
\IEEEauthorblockA{\textit{Renmin University of China} \\
dongjinhao@ruc.edu.cn}
\and
\IEEEauthorblockN{Zan Wang}
\IEEEauthorblockA{\textit{Tianjin University} \\
wangzan@tju.edu.cn}
\and
\IEEEauthorblockN{Qingchao Shen}
\IEEEauthorblockA{\textit{Tianjin University} \\
qingchao@tju.edu.cn}
\and
\IEEEauthorblockN{Junjie Chen}
\IEEEauthorblockA{\textit{Tianjin University} \\
junjiechen@tju.edu.cn}
\and
\IEEEauthorblockN{Wei Lu}
\IEEEauthorblockA{\textit{Renmin University of China} \\
lu-wei@ruc.edu.cn}
\and
\IEEEauthorblockN{Xiaoyong Du}
\IEEEauthorblockA{\textit{Renmin University of China} \\
duyong@ruc.edu.cn}
}

\maketitle

\begin{abstract}
 Question-answering software is becoming increasingly integrated into our daily lives, with prominent examples including Apple Siri and Amazon Alexa. Ensuring the quality of such systems is critical, as incorrect answers could lead to significant harm. Current state-of-the-art testing approaches apply metamorphic relations to existing test datasets, generating test questions based on these relations. However, these methods have two key limitations. First, they often produce unnatural questions that humans are unlikely to ask, reducing the effectiveness of the generated questions in identifying bugs that might occur in real-world scenarios. Second, these questions are generated from pre-existing test datasets, ignoring the broader context and thus limiting the diversity and relevance of the generated questions. 

In this work, we introduce \ourtool, a context-driven question generation approach for testing question-answering systems. Specifically, \ourtool{} extracts entities and relationships from the context to form ground truth answers, and utilizes large language models to generate questions based on these ground truth answers and the surrounding context. We also propose the consistency verification and constraint checking to increase the reliability of \llm{}'s outputs. Experiments conducted on three datasets demonstrate that \ourtool{} outperforms state-of-the-art approaches on  the bug detection capability, the naturalness of the generated questions as well as the coverage of the context. 
% in terms of bug detection, true positive rate, the naturalness of the generated questions \ls{as well as the coverage of context}. 
Moreover, the test cases generated by \ourtool{} reduce error rate when utilized for fine-tuning the QA software under test. 
\end{abstract}


\section{Introduction}
\label{sec:intro}

% \begin{figure}[htb]
%     \centering

%     \begin{subfigure}
%         \includegraphics[width=\linewidth]{figure2-1.png}
%         \centering
%         {\small \textbf{(a)} The generated question contains two irrelevant topics.}
%         \label{fig:example-a}
%     \end{subfigure}
%     \hfill
%     \begin{subfigure}
%         \includegraphics[width=\linewidth]{figure2-2.png}
%         \centering
%         {\small \textbf{(b)} QAQA cannot generate questions covering the highlighted context.}
%         \label{fig:example-b}
%     \end{subfigure}

%     \caption{The deficiencies of questions generated by QAQA}
%     \label{fig:example}
% \end{figure}
\begin{figure}[htb]
    \centering

    \begin{subfigure}[b]{1\linewidth}
        \centering
        \includegraphics[width=\linewidth]{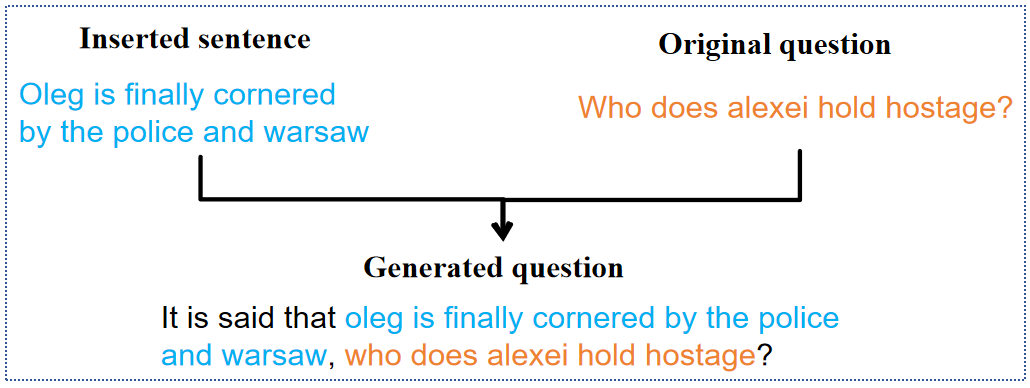}
        \caption{The generated question contains two irrelevant topics.}
        \label{fig:example-a}
    \end{subfigure}
    \hfill
    \begin{subfigure}[b]{1\linewidth}
        \centering
        \includegraphics[width=\linewidth]{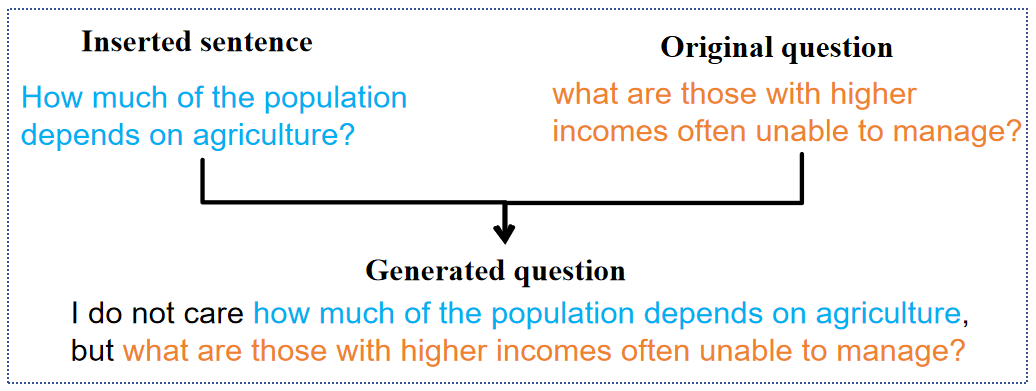}
        \caption{The generated question is not natural}
        \label{fig:example-b}
    \end{subfigure}
    \hfill
    \begin{subfigure}[b]{1\linewidth}
        \centering
        \includegraphics[width=\linewidth]{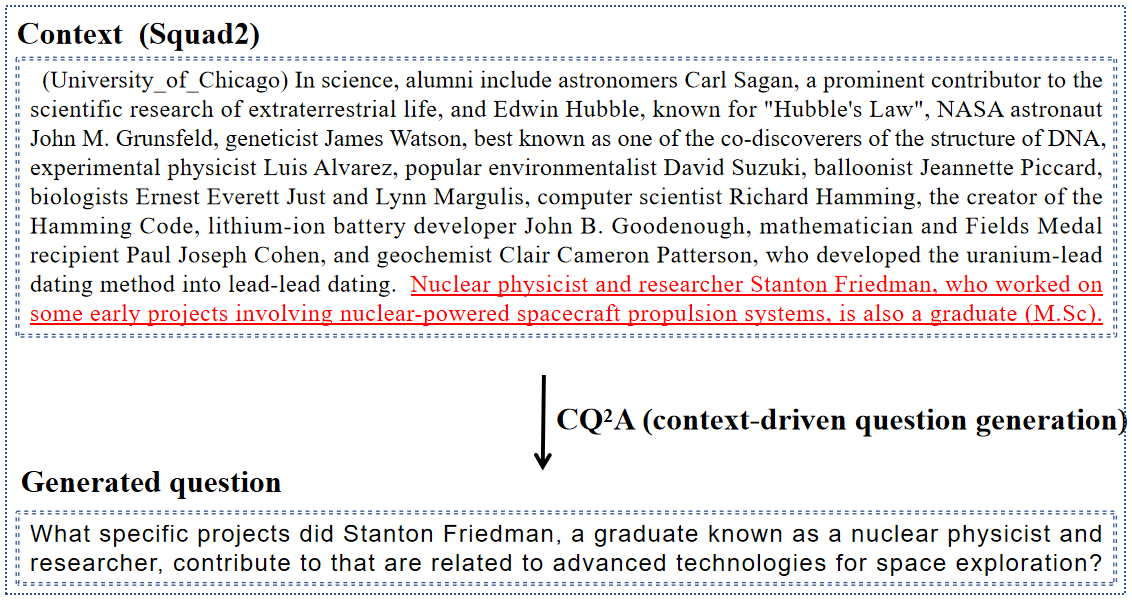}
        \caption{QAQA cannot generate questions covering the highlighted context.}
        \label{fig:example-c}
    \end{subfigure}
    \caption{The deficiencies of questions generated by QAQA.}
    \label{fig:example}
\end{figure}
% \begin{figure}[H]
%     \centering
   
% \end{figure}

% \ls{The application scenarios of QA-system with context.}

Question-answering (QA) systems utilize information retrieval and natural language processing (NLP) techniques to automatically extract and generate answers from various knowledge sources. The primary function of such software is to accept natural language questions posed by users and provide accurate answers based on context, e.g., predefined knowledge bases or reference texts~\cite{zhang2020machine}. With the advancement of artificial intelligence and big data technology, question-answering systems have been widely applied across various fields such as customer service, education and healthcare. Notable examples include Apple Siri~\cite{siri2025}, Amazon Alexa~\cite{alexa}, and Baidu DuerOS~\cite{dueros}. 

% Large language models (LLMs) have recently achieved superior advancement.\jinhao{This sentence lacks coherence with the preceding one.}
% \ls{The necessaries of testing such systems.}
% Although QA software brings significant convenience to our lives, it also contains bugs that can lead to incorrect answers, causing harm to users. 
QA systems are becoming important information acquisition tools in our daily lives, it is thus critical to ensure the correctness of QA systems. Misleading or incorrect answers can lead to significant inconvenience or even substantial losses. 
For instance, in November 2024, Google's AI chatbot Gemini faced criticism after replying with ``please die'' to a student seeking homework help, calling the user ``a stain on the universe''~\cite{chatbot}. In February 2023, Alphabet Inc. experienced a \$100 billion drop in market value after its AI chatbot, Bard, provided inaccurate information in a promotional video~\cite{chatloss}.
% during the U.S. presidential election, people consulting polling station locations often received incorrect or empty addresses~\cite{elect}. 
% \ls{This example is not attractive, no losses caused. }
% Additionally, while many software systems perform excellently on benchmark datasets, they often perform poorly in real-world applications due to discrepancies between the distribution of training data and actual usage scenarios~\cite{miller2020effect,shen2020multiple}. 
% Therefore, there is a need for an testing method that can better simulate daily usage situations to ensure the quality of QA software.\\
% Moreover, some QA systems perform well on benchmark datasets, but there is no guarantee on the performance of real-world application scenarios, as the distribution of training data and actual usage scenarios could be very different~\cite{miller2020effect,shen2020multiple}.
It is thus desirable to test QA systems promptly and efficiently.
% \jinhao{Our technique does not deal with the problem that the distribution of training data and actual usage scenarios are different, this sentence may be removed. The limitations are detailedly introduced in the next two paragraphs, the connections between this sentence and the following paragraphs are not clear} 

% \ls{The challenges of testing such systems. }
% \ls{The shortcomings of existing approaches.}
Testing a question answering software often requires a large number of natural language queries as test cases~\cite{hariri2023unlocking,maulud2021state}. Traditional reference-based testing methods~\cite{longpre2021entity,ribeiro2020beyond} rely primarily on  manually annotated data, which is labor-intensive, limits the ability to detect bugs in a timely manner and suffers from restricted testing adequacy. 
% Annotators are required to design questions and corresponding answers based on a given text. 
% However, this approach has several drawbacks~\cite{chen2021testing}: 1) the manual annotation process is time-consuming, making it difficult to quickly adapt to software updates and iterations; 2) the cost of human labor for annotation is high.

% In the first category of work, researchers extract information for generating questions from pre-established public knowledge bases (e.g., knowledge graphs) and use heuristic rules along with NLP tools~\cite{kamigaito2020syntactically,reimers2019sentence} to construct test cases. 
% \ls{The above two approaches are doing testing after model training?}
% These test cases are then run on the software under test, and the results are compared with the predefined answers.In the second category, researchers follow the principles of metamorphic testing, designing semantically consistent mutation rules to generate test cases. If the answers to the newly generated test cases, when run on the software under test, differ from those of the seed test cases, the software is considered to have a bug. 
Recent testing methods~\cite{chen2021testing,shen2022natural} for QA software follow the principles of metamorphic testing, designing semantically consistent mutation rules to generate test cases.
% QAQA~\cite{}, for instance, propose five metamorphic relations to guide generating new test cases and compare testing results. 
They mainly focuses on modifying existing datasets (such as BoolQ~\cite{clark2019boolq}, SQuAD~\cite{rajpurkar2016squad,rajpurkar2018know}, etc.) by making slight additions, deletions, or synonym substitutions to create new test cases that yield consistent expected answers. However, this kind of approaches have two limitations. First, \textit{the generated test cases often include multiple unrelated topics or lack naturalness, resulting in ineffective bug detection}. For example, Fig~\ref{fig:example-a} and Fig~\ref{fig:example-b} shows the test cases generated by QAQA~\cite{shen2022natural}, 
% ``Nobody cares what university studied `predictable doom,' but why should one not go to jail?'', 
Second, \textit{existing methods primarily focus on the questions in the original question datasets and overlook the abundant information in the contexts, resulting in low testing adequacy.} Since users may ask questions based on the entire context, the test cases generated by existing methods fail to adequately cover the scope of user inquiries, leading to poor testing results. Fig~\ref{fig:example-c} shows a context from the dataset, in which the red-highlighted context information in the last sentence is not covered by any question in its original question dataset. Therefore, existing approach like QAQA~\cite{shen2022natural} and QAASker~\cite{chen2021testing} are unable to generate questions that can cover this context information. 
% such a test case generated by QAQA, 
% \ls{Add a figure of the two counter-examples.}
% \jinhao{Give an example for this limitation}

% \ls{What do we do to tackle the challenges. }
In this work, we propose a novel QA software testing technique, \ourtool, to address the aforementioned limitations. 
% To refine the generated questions so they feel more natural and conversational (Limitation 1), we leverage the latest large language models(LLMs) as tools, using the extracted information as the target for question generation.
Firstly, \ourtool{} employs large language models (LLMs) for test case generation, which is capable of generating human-like questions. 
Secondly, \ourtool{} extract entity and relation information from the whole context, which serve as the ground truth answer and are input to the LLM together with the context for question generation.  
This design enables \ourtool{} to generate human-like questions, with a good coverage on the context. Moreover, given the inherent randomness~\cite{atil2024llm,yadkori2024believebelievellm} and hallucination issues~\cite{ji2023survey} associated with LLMs, we propose a consistency verification and constraint-checking approach to enhance the reliability of \llm{}'s outputs.
% To better cover the scope of user queries (Limitation 2), we utilize NLP tools to extract information from the context of the seed dataset and use it as the standard answer for new test cases, aiming to ensure that the generated questions cover the entire context. During implementation, to improve the generation efficiency of LLMs, we integrate various prompt engineering techniques and strategies to design effective prompts. Moreover, considering the randomness~\cite{atil2024llm,yadkori2024believe} and hallucination issues~\cite{ji2023survey} inherent in LLMs, we have also designed a series of filtering criteria to rigorously select the generated output, ensuring the quality of the results.

For test result comparison, the questions posed to QA systems are often open-ended, making exact matches an insufficient method for evaluating answer correctness. Moreover, ignoring context can lead to false positives, as two semantically inconsistent answers could actually refer to the same concept within a given context. To address this, our approach employs a two-stage process, combining semantic similarity analysis with LLM-assisted judgment as the test oracle, to assess whether an answer aligns with the ground truth. The robust language comprehension abilities of LLMs enhance answer comparing accuracy and significantly reduce false positives in bug detection. 

% \jinhao{Comment: I rewrite this paragraph, you said the previous pre-trained models have limited abilities so you use LLMs, but LLMs are still pre-trained models. I think we do not mention there exist limitations and we diretly say what we do}
% In addition, existing methods often rely on pre-trained models to compute semantic similarity between the answers on seed test cases and new generated test cases, or on semantic judgments of individual words~\cite{chen2021testing}. These detection methods may lead to errors in judgment for two main reasons: First, due to the performance limitations of the models, they may fail to accurately determine whether the semantics of the two answers align, potentially resulting in cases where the semantic similarity is low, but the actual meaning is consistent; Second, the lack of consideration for the context can lead to situations where two answers, despite being semantically inconsistent, refer to the same thing in the context, resulting in false positives. Therefore, in our method, we combine semantic similarity with large model-assisted judgment as the test oracle. This dual-layered approach allows for more accurate identification of bugs while reducing the incidence of false bug reports.\\

We conduct an experimental study to investigate the effectiveness of \ourtool. In particular, we choose the latest state-of-the-art testing methods QAQA~\cite{chen2021testing} and QAAskeR-plus~\cite{xie2023qaasker+} as the baselines.  
Our experiments use three datasets, i.e., BoolQ~\cite{clark2019boolq}, SQuAD2~\cite{rajpurkar2016squad,rajpurkar2018know}, and NarrativeQA ~\cite{kovcisky2018narrativeqa}, which are commonly adopted by existing approaches~\cite{shen2022natural,chen2021testing} to evaluate the testing effectiveness. 
We select the state-of-the-art QA software UnifiedQA-v2~\cite{khashabi2020unifiedqa,khashabi2022unifiedqa} as the software under test. 
Our experimental evaluation focuses on the following aspects: (1) the ability of our method on exposing bugs, (2) the naturalness and context coverage of the generated test cases, (3) the effectiveness of the filtering process, and (4) the impact of the generated test cases on model repair. The results demonstrate that \ourtool{} can reveal more bugs, with 45.17\% higher true positive rates than that of QAQA and 22.84\% than that of QAAskeR-plus. It can generate more natural test cases. The test cases generated by \ourtool{} have an average of 43.7\% higher context coverage than that of QAQA and 24.7\% higher context coverage than that of QAAskeR-plus. Moreover, the test cases show effectiveness in repairing the QA models, with an average of 30.2\% error rate decrease.  
% In terms of quality evaluation of the question in three dimensions, our method completely outperforms QAQA. In addition, the filtering process is quite effective in promoting the quality of the the test case. Furthermore, the test cases generated by our method also contribute to the effective repair of the UnifiedQA model.
% \ls{summary of contributions.}

Our work makes the following contributions.
\begin{itemize}
     \item \textbf{Technique}: We have designed a novel QA software testing method, \ourtool, which features a context-driven test case generation method employing large language models, consistency verification to ensure the quality of the generated questions, and a test oracle that combines semantic similarity with large language model-assisted judgment to compare answer consistency. 
     % \ourtool{} addressing the shortcomings of existing approaches that focus only on isolated semantic defect detection.
     % \jinhao{Split this paragraph to several contribution points, each point should be short. Add a descriptive word to each point, similar to the following}
    \item \textbf{A comprehensive experiment}: We have conducted comprehensive experiments to evaluate our method's effectiveness in bug detection, test case naturalness, context coverage, filtering effectiveness, and the impact of generated test cases on model repair. The evaluation results show that \ourtool{} outperforms the compared baselines on all those aspects. 
    \item \textbf{A replication package}: We have packaged our method and released the code, along with our experimental data and results, on our project homepage~\cite{CQA} to facilitate replication studies and future research.
\end{itemize}

\begin{figure*}[t]
  \centering
  \includegraphics[width=0.88\linewidth]{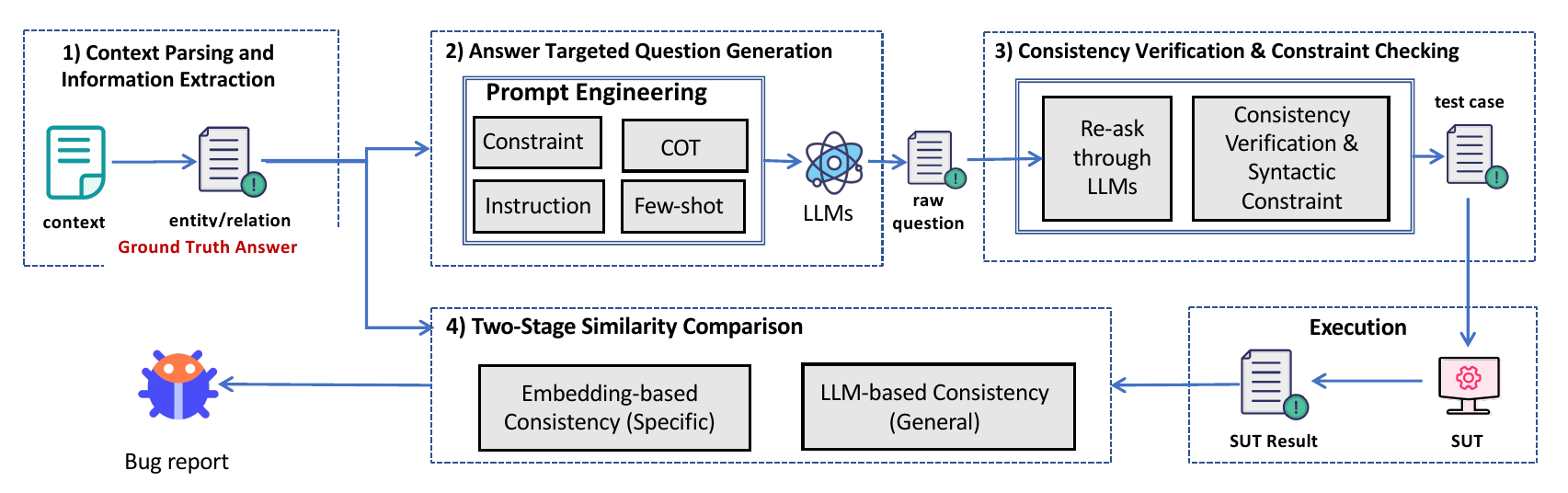} %1.png是图片文件的相对路径
  \caption{The overview of \ourtool{}} %caption是图片的标题
  \label{fig:overview} %此处的label相当于一个图片的专属标志，目的是方便上下文的引用
\end{figure*}

\section{Preliminary}
\subsection{Question-Answering Software}
% \subsubsection{Closed-World Question-Answering Software}
Question-answering (QA) software has long been a popular research topic, frequently appearing in fields such as healthcare~\cite{athenikos2010biomedical,budler2023review}, e-Commerce~\cite{gao2021meaningful,gao2019product} and smart home~\cite{nandy2021question} applications. Based on the format of the information required to answer questions, QA software can be classified into open-world QA and closed-world QA software~\cite{chen2021testing}. The former requires users to input a question, and it will provide an answer based on a pre-established open knowledge base. Notably, large language models (LLMs) can also be regarded as open-world QA systems~\cite{keluskar2024llms}. The latter requires users to input both the question and the context needed to answer the question, and the software will respond only based on the context. 

Closed-world QA software is cost-effective and can be easily fine-tuned to specific domains~\cite{zaib2022conversational}. Moreover, in domain-specific scenarios, it often demonstrates higher accuracy compared to open-world QA systems~\cite{kia2022adaptable,soto2024rag}. Consequently, it is widely utilized in applications which has high requirement on the answer correctness, such as medicine and legal applications~\cite{do2024r2gqa,sun2023generating}. Given the wide application of closed-world QA software and its critical application scenarios, a number of recent  research works~\cite{chen2021testing,shen2022natural} have focused on testing closed-world QA software. 

\subsection{Large Language Models}
% \ls{We'd better remove this part as we did not use LLM as a testing target. We can move discussions on the hallucination problems to the method part.}
% Large language models (LLMs), such as GPT and Llama, have revolutionized the field of natural language processing. These models are trained on vast datasets using deep learning techniques, particularly the Transformer architecture, to understand and generate human-like text. With billions or even trillions of parameters, LLMs are capable of capturing complex language patterns, making them effective across various NLP tasks such as text generation, translation, summarization, and more.

% One of the most prominent applications of LLMs is text generation, where the model produces coherent, contextually relevant text based on an input prompt. Currently, many researchers focus on how to efficiently generate text using large models, and due to issues such as the randomness~\cite{atil2024llm,yadkori2024believe} and hallucination ~\cite{ji2023survey} in text generation, a significant body of work has also concentrated on evaluating and filtering the results generated by large models.

Large language models (LLMs), such as GPT~\cite{chatgpt} and LLaMA~\cite{touvron2023llama}, have revolutionized natural language processing. These models, built on the Transformer architecture and trained on extensive datasets, are capable of understanding and generating text. With billions or even trillions of parameters, LLMs capture complex language patterns, excelling in tasks like text generation, translation, summarization, and more.
Text generation is among the most prominent applications of large language models (LLMs), enabling the production of coherent and contextually appropriate text based on input prompts. Extensive research has focused on enhancing the effectiveness and quality of text generated by LLMs across a variety of tasks. In this study, we leverage the strong text generation capabilities of LLMs to generate questions given a specific context and answer. However, challenges such as output variability~\cite{atil2024llm,yadkori2024believebelievellm} and hallucinations~\cite{ji2023survey} pose significant threats to the reliability of LLM-generated results. In our work, hallucinations may lead to generated questions that introduce information not present in the source context or violate the constraints specified in the prompt.  We further introduce a consistency verification technique designed to identify and filter out false positives among the questions generated by LLMs, improving the quality and reliability of LLM-generated outputs.

\section{Method}
\label{sec:method}

% \textit{existing methods primarily focus on the questions in the datasets and overlook the abundant information in the contexts, resulting in low testing adequacy.}

% \textit{the generated test cases are often unnatural, resulting in ineffective bug detection}

% Existing closed-world QA testing techniques primarily focus on modifying existing datasets, overlooking the wealth of contextual information. This limitation reduces testing adequacy, while the transformation rules often result in unnatural test cases.

In this paper, we introduce a context-driven, answer-targeted testing technique, \ourtool{}. Distinct from existing test generation approaches, \ourtool{} extracts entities/relationships from the context to serve as answers, and subsequently generates corresponding questions. This strategy enables the generation of more diverse test cases and ensures broader coverage of the target domain. Moreover, in contrast to the rigid concatenation approach employed by \qaqa{}~\cite{shen2022natural}, our method leverages large language models (\llm{}) to produce questions that are notably more coherent and natural. As a result, \ourtool{} significantly enhances both the completeness and naturalness of testing for QA systems.

An overview of \ourtool{} is presented in Fig.~\ref{fig:overview}. To achieve improved completeness and naturalness, \ourtool{} incorporates four key components: (1) context parsing and information extraction to obtain ground truth answers from the context (Section~\ref{sec:context}), (2) answer-targeted question generation (Section~\ref{sec:answer}) to generate questions with the given context and ground truth answers, (3) consistency validation and constraint checking to enhance the reliability of \llm{}'s generation results (Section~\ref{sec:filter}), and (4) two-stage similarity checking to improve comparison accuracy (Section~\ref{sec:comparison}). 

\begin{table*}[tb]
    \centering
    \caption{Prompt Designs for Test Generation}
    \label{tab:CQ2A_qg_design}
    \renewcommand{\arraystretch}{1.1}  % Adjust row spacing
    \begin{tabularx}{\textwidth}{>{\raggedright\arraybackslash}p{3cm}>{\raggedright\arraybackslash}X}
        \toprule
        \textbf{Prompt Type} & \textbf{Prompt Content} \\
        \midrule
        Relation Extraction Prompt & 
        % \setlength{\baselineskip}{1.3\baselineskip} Extract relationships between noun entities in the text and provide reasonable explanations.\newline
        % \textbf{Rules:} \newline
        % 1. Ensure extraction results are unambiguous. If a pronoun (e.g. he, she) appears in the triple, replace it with the corresponding word.\newline
        % 2. Reasoning about implicit entity relationships based on the text is allowed. Output as many relation triples as possible.\newline
        % 3. The extracted relation must be a content verb or noun, not a linking verb (e.g., is, are, have). Triples should be separated by commas, and the two entities must not be the same.\newline
        % \textbf{Example:} Text: "Lucius Harney becomes Mr. Royall's boarder." \newline
        % Relation: [Lucius Harney, becomes, Mr. Royall's boarder.] \\

\setlength{\baselineskip}{1.3\baselineskip} Extract relationships between noun entities in the text and provide clear explanations.\newline
        \textbf{Rules:} \newline
        1. Ensure that extraction results are explicit and unambiguous. If a pronoun (e.g., he, she) appears in a triple, replace it with the corresponding noun.\newline
        2. Reasoning about implicit relationships between entities based on the text is allowed. Extract as many relation triples as possible.\newline
        3. The extracted relation should be a content verb or noun, rather than a linking verb (e.g., is, are, have). Separate each triple with a comma, and ensure the two entities in a triple are not identical.\newline
        \textbf{Example:} Text: Lucius Harney becomes Mr. Royall's boarder. \newline
        Relation: [Lucius Harney, becomes, Mr. Royall's boarder.] \\

        % \hline
        % Entity-Based Question Generation Prompt & \setlength{\baselineskip}{1.3\baselineskip} Design at least five questions centered on a specific word or phrase in the sentence. Ensure the answer is that word or phrase, and provide a justification for each question.\newline
        % \textbf{Requirements:} \newline
        % 1. Questions should be specific enough to avoid ambiguity and ensure a unique answer.\newline
        % 2. To increase complexity, additional contextual information may be incorporated. Questions must be longer than 25 characters.\newline
        % 3. Questions should be natural and human-like in phrasing.\newline
        % \textbf{Example Format:} \newline
        % (1) Question: [...] Explanation: [...] \newline
        % (2) ... (3) ... \\
        % \hline
        % Relation-Based Question Generation Prompt & \setlength{\baselineskip}{1.3\baselineskip} Design at least ten questions based on relation triples [Entity1, Relation, Entity2], and provide explanations.\newline
        % \textbf{Requirements:} \newline
        % 1. The answer must be the relation in the triple.\newline
        % 2. The explanation must be strictly based on the question content.\newline
        % 3. The question must include Entity1 and Entity2, but must not directly contain the relation.\newline
        % 4. The question must be specific enough to ensure a unique answer.\newline
        % \textbf{Example Format:} \newline
        % (1) Question: [...] Explanation: [...] \newline
        % (2) ... (3) ... \\
\hline
Entity-Based Question Generation Prompt & \setlength{\baselineskip}{1.3\baselineskip} Design at least five questions focused on a specific word or phrase within the sentence. Ensure that the answer to each question is exactly that word or phrase, and provide a clear explanation for each question.\newline
\textbf{Requirements:} \newline
1. Each question should be precise enough to avoid ambiguity and guarantee a single, unique answer.\newline
2. To enhance complexity, you may incorporate additional contextual details. Each question must exceed 25 characters in length.\newline
3. The questions should be phrased naturally, as if posed by a human.\newline
\textbf{Example Format:} \newline
Question: [...] Explanation: [...] \\
\hline
Relation-Based Question Generation Prompt & \setlength{\baselineskip}{1.3\baselineskip} Design at least ten questions based on relation triples in the format [Entity1, Relation, Entity2], and provide explanations for each.\newline
\textbf{Requirements:} \newline
1. The answer must be the relation described in the triple.\newline
2. The explanation should be strictly derived from the content of the question.\newline
3. Each question must mention both Entity1 and Entity2, but should not directly state the relation.\newline
4. The question should be sufficiently specific to ensure a unique answer.\newline
\textbf{Example Format:} \newline
Question: [...] Explanation: [...]\\

        \bottomrule
    \end{tabularx}
\end{table*}

\subsection{Context Parsing and Information Extraction}
\label{sec:context}

A closed-world QA system must answer questions based on the provided context. To sufficiently test such a system, the generated questions should comprehensively cover the information in the context. To achieve this, we first parse the context and extract its key information. This extracted information serves as the foundation for test generation, enabling us to create tests that effectively cover the context and improve the overall completeness of the testing process.

% This section leverages the context information from the seed dataset, using LLms as tools to generate high-quality test cases. It includes the following three steps: Information Extraction, Question Generation Based on LLms, and Result Filtering.
% \subsubsection{Information Extraction}
% This step extracts information from the context to serve as the answer for generating questions. 

To effectively utilize contextual information and maintain better control over the question generation in subsequent components, we predefine two categories of information: entity information (nouns, verbs, noun phrases, verb phrases) and relationship information (relationship triples). The context is parsed to extract these two types of information.

\subsubsection{Entity Extraction}
For entity information, we analyze the text at the sentence level using Stanza~\cite{stanza}, a widely used NLP toolkit developed by Stanford University~\cite{lan2023part}. Firstly, Stanza is employed for part-of-speech tagging and syntactic analysis to extract preliminary results.
After the initial parsing, we extract information such as noun phrases from the structured data generated by the system, which contains entity-related information. However, these preliminary results may include redundant or inaccurate content. To address this, we apply a set of manually defined filtering rules to refine the initial outputs. These rules are designed based on the types and characteristics of entities, aiming to improve the accuracy and validity of the extracted information. The filtering rules are as follows:
% Subsequently, syntactic dependency relations are leveraged to apply filtering rules that refine the results. The filtered entity information is stored as the output. 

\begin{itemize}
    \item \textbf{Phrases:}
    \begin{itemize}
        \item The length of the phrase must be at least two words.
        \item If a phrase is found to contain clause elements according to syntactic analysis, these clause components are removed.
        \item Duplicate phrases are eliminated.
    \end{itemize}

    \item \textbf{Words:}
    \begin{itemize}
        \item Nouns must not include pronouns; verbs must exclude non-lexical verbs and cannot function as modifiers or infinitives in the dependency relations.
        \item Words extracted as part of phrases in the same sentence are excluded to avoid redundancy.
    \end{itemize}
\end{itemize}
\subsubsection{Relationship Extraction}
% For relationship information, the objective is to extract relationship triples in the format [entity1, relation, entity2], where the relation serves as the standard output. Given the limitations of existing relationship extraction tools, we employ the state-of-the-art large language model ChatGPT~\cite{chatgpt} for this task. By combining few-shot learning~\cite{wang2020generalizing} with chain-of-thought prompting~\cite{wei2022chain}, we design specialized prompts for relationship extraction. Using these prompts, ChatGPT generates initial relationship triples, which are further refined through heuristic filtering rules to produce the final results. The filtering rules are as follows:

% Both entities must appear in the explanation and the context, and they must be either nouns or noun phrases.
% The relation must be a lexical verb.
% This methodology ensures the systematic extraction of high-quality entity and relationship information while leveraging advanced NLP and LLM techniques.

The goal of extracting relationship information is to identify relationship triples in the format [entity1, relation, entity2], where the entities act as the components of the query, and the relation represents the answer in the test. Due to the limitations of existing relationship extraction tools, we leverage the state-of-the-art LLM ChatGPT~\cite{chatgpt} for this task. By integrating few-shot learning~\cite{wang2020generalizing} with chain-of-thought prompting~\cite{wei2022chain}, we develop specialized prompts tailored for relationship extraction. The prompt for extracting relation is shown in the first row of Table~\ref{tab:CQ2A_qg_design}. Using this prompt, ChatGPT extracts initial relationship triples. To mitigate hallucinations and ensure the accuracy of the extracted relations, we introduce a set of filtering rules. These rules are as follows: (1) Both entities must appear in the context and the explanation generated by ChatGPT. This ensures that the entities of the relation genuinely exist within the context, making the relation reasonable and reducing the occurrence of hallucinations. (2) The entities should be either nouns or noun phrases, while the relation should be a lexical verb.

% In summary, This methodology ensures the systematic extraction of high-quality entity and relationship information by leveraging advanced NLP and LLM techniques.

\subsection{Answer-Targeted Question Generation}
\label{sec:answer}
% In this step, questions are generated based on the answers to create test cases. To ensure the generated questions are more natural and diverse, thus enabling to detect more bugs, we choose ChatGPT as the tool for question generation. Specifically, we input the information obtained in step one as answers, combined with the sentence and context, into the ChatGPT. The model then generates a question based on the answer. To generate questions more efficiently, we use various prompt engineering techniques. Hence the prompt consists of four parts: `instructions',`constraints',`multi-examples' and `chain-of-thought'.
\subsubsection{Reverse Order for Test Case Generation}
% Existing techniques leverage some semantically consistent mutation rules to construct new tests (e.g., concatenating two unrelated sentences), which will produce unnatural tests and makes the testing ineffective. In addition, how to generate the oracle (i.e., ground truth outputs) is challenging when generating tests, since we cannot verify whether the generated oracles are correct. 

% To address these two challenges, we propose an answer-targeted test generation approach. In reverse order, we first select the answers and then generate the corresponding questions. This method ensures that the answers and questions are naturally paired, resulting in coherent and realistic questions. Furthermore, by starting with the answers, we effectively solve the issue of lacking oracles.

Existing methods construct new test cases using semantically consistent transformation rules (e.g., synonym substitution, word order changes). However, these approaches often generate unnatural test cases, reducing the effectiveness of the evaluation. Moreover, since it is difficult to directly verify whether the answers provided by the QA system are correct during testing, defining reliable reference results poses a significant challenge.

To effectively address the above two issues, \ourtool{} proposes an answer-oriented test case generation method. Unlike traditional approaches, \ourtool{} adopts a reverse generation strategy—first determining the answer, then generating corresponding questions based on it. By explicitly specifying the test objective—the predefined answer—this method ensures that the generated questions are more naturally aligned with their answers, thereby enhancing the coherence and tight coupling between them. Since the test cases are generated from known answers and utilize LLMs, the resulting questions better reflect real users’ natural questioning patterns.

Moreover, the answer-oriented approach effectively addresses the lack of reference results in traditional methods. Since the answer is predefined, we can directly detect system flaws by comparing the response generated by the system under test with the expected answer. The predefined answer can also assist in evaluating the quality of the generated questions. By starting from the answer, the relationship between questions and answers is fully guaranteed, which improves the quality of the generated test cases and provides a more reliable foundation for subsequent QA system evaluation.

\begin{table*}[tb]
    \centering
    \caption{Prompt Designs for Re-asking and Consistency Evaluation}
    \label{tab:CQ2A_evaluation_design}
    \renewcommand{\arraystretch}{2}
    \begin{tabularx}{\textwidth}{>{\raggedright\arraybackslash}p{2.5cm}>{\raggedright\arraybackslash}X}
        \toprule
        \textbf{Prompt Type} & \textbf{Prompt Content} \\
        \midrule
        \setlength{\baselineskip}{1.5\baselineskip}
        Re-asking Prompt &
        \setlength{\baselineskip}{1.5\baselineskip}
        Given the provided \textit{context} and \textit{question}, output the answer. \newline
        \textbf{Requirements:} \newline
        1. Output strictly the answer, with no additional content. \newline
        2. Ensure that the answer is precise and unambiguous. \\
        \hline
        \setlength{\baselineskip}{1.5\baselineskip}
        LLM-Assisted Consistency Evaluation Prompt &
        \setlength{\baselineskip}{1.5\baselineskip}
        Given the \textit{context} and \textit{question}, evaluate the consistency between \textit{gold\_answer} and \textit{sut\_answer} (the answer generated by the system under test). \newline
        \textbf{Evaluation Criteria:} \newline
        1. The answers exhibit high semantic similarity. \newline
        2. Both answers refer to the same information. \newline
        3. The content of \textit{gold\_answer} is contained in \textit{sut\_answer}, and the entirety of \textit{sut\_answer} appears explicitly in the context. \newline
        \textbf{Scoring Guidelines:} \newline
        - If the above criteria are met, the answers are considered consistent and should receive a high score. \newline
        - The score ranges from 0 (inconsistent) to 100 (fully consistent). \newline
        - Each scoring result must include an explanation in the following format: \newline
        \textbf{``[score] \textbackslash n [explanation]''} \\
        \bottomrule
    \end{tabularx}
    \vspace{-2mm}
\end{table*}

\subsubsection{Extracted Entities and Relationships Serve as Answers}
Specifically, we leverage the entities and relationships extracted in Section~\ref{sec:context} as answers to the tests and generate corresponding questions based on these answers. \llm{} demonstrates a strong capability for understanding and generating text. To ensure that the generated questions are more natural and diverse—thereby enhancing the detection of potential bugs—we employ \llm{} (\chatgpt{} in this paper) for question generation. The entities and relationships extracted in the previous step are provided as answers, along with the sentence and context, to \llm{}, which then generates questions based on the given answers.

\vspace{1mm}
\noindent\textbf{Entity-Based Question Generation. }
To enhance the effectiveness of question generation based on the extracted entities, \ourtool{} explicitly incorporates a set of constraint conditions into its prompt design. These constraints are intended to ensure that the generated questions are both precise and unambiguous, thereby preventing the generation of incorrect questions caused by semantic vagueness or ambiguity. For example, the prompts require the questions to exhibit clear referential direction, ensuring that each question corresponds to a unique and clearly identifiable answer. This significantly improves the accuracy of the generated test cases. In addition, \ourtool{} pays attention to the length and content of the questions. Overly simplistic questions contribute little to identifying defects in QA systems. Therefore, to ensure the quality of the generated questions, \ourtool{} enforces length constraints, thereby reducing the production of low-value test cases.
To further enhance the language model’s understanding during the generation process, \ourtool{} incorporates the concept of Chain-of-Thought (CoT) reasoning. Under this approach, the language model is required not only to generate a question but also to provide an accompanying explanation—justifying why the predefined entity serves as the correct answer to the generated question. This allows the model to better understand the underlying logical relationships and adhere more closely to coherent structures during question generation, thereby reducing ambiguity and inconsistency. This method not only improves the quality of the generated questions but also enables large language models to produce more accurate outputs in complex question generation tasks. The complete prompt for entity test generation is illustrated in the second row of Table~\ref{tab:CQ2A_qg_design}.

\vspace{1mm}
\noindent\textbf{Relation-Based Question Generation. }
The extracted relational information is presented in the format [Entity1, Relation, Entity2]. In this process, \ourtool{} requires the large language model to generate corresponding questions with the relation in the triple as the answer. Specifically, the model must formulate a question based on the context, ensuring that the relation is treated as the answer. Additionally, both entities must be included in the question to guarantee that the generated question accurately reflects the relationship between the entities within the text. However, experimental results have demonstrated that this task presents a challenge for large language models. While these models excel in natural language generation, they often encounter difficulties in simultaneously understanding both the triple and the surrounding context, as well as in formulating questions that effectively address the given relation.

To address this issue, \ourtool{} proposes several optimization strategies. First, similar to how entity information is handled, this section also adopts a chain-of-thought reasoning approach, requiring the language model to articulate the logical basis and reasoning process behind the generated question. In addition, \ourtool{} employs a few-shot learning strategy, in which a set of carefully designed examples—including sample question-generation results and corresponding reasoning explanations—are embedded in the prompt. These demonstrations guide the model to learn how to extract the relational information from a given triple and accurately convert it into a question. The specific prompt of relation test generation is shown in the third row of Table \ref{tab:CQ2A_qg_design}. Since relation-based question generation is more challenging and may result in a higher proportion of invalid questions, we instruct the large language model to generate 10 questions for the relation-based setting and 5 questions for the entity-based setting.

% In this step, questions are generated based on the answers to create test cases. To ensure the generated questions are more natural and diverse, thereby enabling the detection of more bugs, we select ChatGPT as the tool for question generation. Specifically, the information obtained in step one is input as answers, combined with the sentence and context, into ChatGPT. The model then generates a question based on the provided answer. To improve the efficiency of question generation, we utilize various prompt engineering techniques. As a result, the prompt is designed with four components: instructions, constraints, multi-examples, and chain-of-thought.

% \subsubsection{Result Filtering}
\subsection{Consistency Verification and Constraint Checking to Enhance Reliability}
\label{sec:filter}
In practical applications, large language models often face the problem of hallucination~\cite{atil2024llm,yadkori2024believebelievellm,ji2023survey}, in which the generated content is grammatically correct and fluent but is not accurate in terms of meaning or facts. This problem can usually be divided into two types: factual hallucination and faithfulness hallucination. Factual hallucination happens when the generated content goes against real-world facts, while faithfulness hallucination happens when the output misinterprets the input information and does not accurately show the intended meaning. In the context of CQ2A, where large language models generate questions based on given answers, the main issue is faithfulness hallucination—meaning that the generated questions may not match the answers correctly. To address this problem, \ourtool{} includes a result filtering module that uses constraint checking and consistency verification to effectively detect and remove issues caused by model hallucinations. The details of the result filtering process are described below.

\subsubsection{Constraint Checking.}
We define the constraints from the following perspectives.
\begin{itemize}
    \item \textbf{Lexical Validity}: 
    To ensure the generated questions are semantically clear and rich in meaning, certain specific words—such as personal pronouns (e.g., ``he'', ``it''), demonstrative pronouns (e.g., ``this'', ``that''), and non-lexical verbs (e.g., ``is'', ``are'')—are restricted from being used as answers. These constraints are designed to eliminate potentially ambiguous vocabulary, ensuring that each question clearly conveys its intended meaning.
    % To ensure that the generated questions are both semantically clear and meaningful, we restrict the use of certain words—such as pronouns or non-lexical verbs—as answers. These constraints are intended to eliminate potentially ambiguous vocabulary, thereby ensuring that each question precisely conveys its intended meaning.
    \item \textbf{Completeness}: When generating a question where the answer is an entity, the entity must also appear in the explanation provided by the large language model. This requirement guarantees that the extracted entity information is faithfully preserved in the generated question. Similarly, for relation-based answers, the corresponding triple should be present in the model’s explanation to ensure the relational information is fully  conveyed.

    \item \textbf{Avoiding Information Leakage}: To prevent information leakage, the answer itself must not appear in the generated question. This requirement preserves the integrity of the test and avoids situations where including the answer in the question could result in invalid or misleading evaluation outcomes.
\end{itemize}

% Heuristic rules are applied to filter the LLM's generated outputs. These rules are defined as follows. Firstly, the answer should not be a pronoun or a non-lexical verb. In addition, the answer should not appear within the question itself. Furthermore,  for entity-based answers, the entity information must appear in both the original sentence and its context; for relationship-based answers, all triples from the explanation must be included.

\subsubsection{Consistency Verification.} 

% To verify the correctness of the questions generated by \llm{}, we propose a consistency verification technique. Starting with the reference answer $ans_{ref}$ extracted from the context, we prompt \llm{} to generate a corresponding question $ques$. Subsequently, the model is asked to re-answer the generated question $ques$ based on the same provided context. The newly generated answer is denoted as $ans_{new}$. To assess alignment between $ans_{ref}$ and $ans_{new}$, we compute their semantic similarity using \simcse{}~\cite{gao2021simcse}, a tool leveraging contrastive learning to enhance sentence embeddings. Only results with a similarity score exceeding 0.75 are retained.
% In addition, to mitigate model bias, a majority-vote mechanism is integrated into the process of obtaining $ans_{new}$. Specifically, five candidate answers are generated, and the most frequent response is selected as the final answer.

% The previous rules can only provide coarse-grained constraints on question generation, but cannot guarantee its correctness, that is, the question and the answer are matched. 
% To make the generated question more reliable, \ourtool{} designs a consistency verification approach. We use the large language model to answer each candidate question again. The resulting answer is then compared to the reference (gold) answer. If the two answers are consistent and the question satisfies the previously defined constraints, the question is deemed valid. Otherwise, it is considered invalid and should be discarded.

The aforementioned rules provide only coarse-grained constraints for question generation and cannot ensure the correctness or alignment between the generated question and its answer. To enhance the reliability of question generation, \ourtool{} introduces a consistency verification mechanism. Specifically, each candidate question is re-answered by the large language model, and the resulting answer is compared with the reference (gold) answer. If the two answers are consistent and the question also satisfies the previously defined constraints, the question is considered valid. Otherwise, it is deemed invalid and is discarded. The implementation consists of two main steps:

\vspace{1mm}
\noindent\textbf{Step 1: Re-asking the generated question. }
\ourtool{} prompts the large language model to perform the ``re-asking'' task, instructing it to generate an answer based on the provided context and question. The detailed prompt format is provided in Table~\ref{tab:CQ2A_evaluation_design}.
To mitigate potential hallucinations in the model’s output, \ourtool{} employs a majority voting mechanism during answer generation. Specifically, the large language model responds to the same prompt five times, after which the most frequently occurring answer is selected as the final result. This approach helps to reduce randomness and instability in the model’s responses, thereby enhancing the reliability of the generated answers.

\vspace{1mm}
\noindent\textbf{Step 2: Answer consistency comparison.}
Once the question and corresponding answer have been generated, \ourtool{} compares the answer produced by the model with the predefined reference (gold) answer. To accomplish this, \ourtool{} utilizes the pre-trained SimCSE model to embed both answers into vector representations, followed by the calculation of their cosine similarity.
Following existing works~\cite{shen2022natural,xie2023qaasker+}, \ourtool{} sets a similarity threshold of 0.75: if the cosine similarity between the two answers is greater than or equal to this threshold, the answers are deemed consistent and the corresponding question is considered valid under the established constraints. If the similarity falls below the threshold, the question is regarded as invalid and is discarded.

By combining consistency verification with constraint checking, we improve the reliability of questions generated by \llm{}, effectively filtering out incorrect questions arising from fidelity hallucinations. Consequently, our method yields a set of high-quality, filtered test cases (question–answer pairs). %The specific format of the re-asking prompt is shown in Table \ref{tab:CQ2A_evaluation_design}.

% \subsection{Oracle Construction}
% \label{sec:oracle}
% This section involves running the test cases generated in the previous part on the software under test to obtain the software's answers. In our experiment, we use the state-of-the-art QA software, UnifiedQA, as the software under test. Therefore, we combine the questions obtained in the previous part with their respective context, modify them to match the format required by the software, and then input them into UnifiedQA to obtain the oracle results.
\subsection{Similarity Checking to Improve Comparison Accuracy}
\label{sec:comparison}

This section aims to determine whether the test cases can effectively expose bugs in the QA software under evaluation. Specifically, the approach involves comparing the standard (reference) answer provided in each test case with the response generated by the QA system. A defect is identified when the two answers are found to be inconsistent.

To assess answer consistency, the state-of-the-art QAQA technique primarily relies on embedding models (e.g., BERT) to compute the semantic similarity between the answers. If the similarity score exceeds a predefined threshold, the answers are considered consistent; otherwise, they are deemed inconsistent. However, embedding models have inherent limitations due to their dependence on fixed-length vector representations and their lack of explicit contextual reasoning. As a result, these models may fail to recognize semantically equivalent answers that differ structurally, or those requiring interpretation based on broader context. In contrast, large language models are capable of more sophisticated reasoning and nuanced comparison, making them a valuable complement to embedding-based methods for more accurate answer consistency evaluation.

To improve comparison accuracy, we propose a two-stage similarity checking method that integrates both embedding models and LLMs. In the first stage, we employ \simcse{}\cite{gao2021simcse}, a contrastive learning-based approach that enhances sentence embeddings, to calculate the semantic similarity between answers. If the similarity score falls below the predefined threshold, we further evaluate the answers using \llm{} to account for contextual variability. In the second stage, we utilize \chatgpt{} to assess the consistency of answers within the given context. The specific prompt format used for consistency comparison is illustrated in the second row of Table~\ref{tab:CQ2A_evaluation_design}. By leveraging the advanced reasoning capabilities of LLMs, this approach improves the accuracy of defect detection while reducing false positives caused by linguistic variations.

A defect is confirmed only when both the embedding model and the LLM judge the answers to be inconsistent.
\section{Experiment}
\label{sec:exp}

\subsection{Experiment Setup}
\subsubsection{Software under Test}
We use UnifiedQA~\cite{khashabi2020unifiedqa,khashabi2022unifiedqa} as the software under test, as it is one of the most advanced QA models and is frequently used in existing studies.
UnifiedQA is a QA language model based on the T5 architecture and can handle multiple question formats. We use the latest version, UnifiedQA-V2, whose pre-trained model has been trained on 20 different QA datasets. The developers have provided several sizes of pre-trained models, and in this experiment, we select UnifiedQA-v2-large, which has stronger ability than small and base versions, as the software under test.
% \usepackage{booktabs}

% \begin{tabular}{ccc} 
% \toprule %表格顶部粗横线
% title1 & title2 & title3 \\
% \midrule % 表格中间细横线
% 1 & 2 & 3 \\
% 4 & 5 & 6 \\
% \bottomrule % 表格底部粗横线
% \end{tabular}

% \begin{table}[h]
% \centering
% \caption*{Table1: Datasets}
% \begin{tabular}{@{}lccccc@{}}
% \toprule
% \textbf{Dataset}  & \textbf{Train set size} & \textbf{Test set size} \\ \midrule
% BoolQ                 & 9,427                     & 3,270                 \\
% SQuAD2               & 130,319                   & 11,873                \\
% NarrativeQA       & 32,747                    & 3,461                 \\ \bottomrule
% \end{tabular}
% \label{tab:dataset}
% \end{table}

\begin{table}[tb]
    \centering
    \caption{QA Datasets}
    \label{tab:dataset}
        % \begin{adjustbox}{width=0.5\textwidth}
    \begin{tabular}{lcc}\toprule
\textbf{Dataset}  & \textbf{Training set size} & \textbf{Test set size} \\ \midrule
BoolQ                      & 9,427                      & 3,270                 \\ 
SQuAD2                 & 130,319                    & 11,873                \\ 
NarrativeQA           & 32,747                     & 3,461                 \\ \bottomrule
\end{tabular}
% \end{adjustbox}
\end{table}
\subsubsection{Datasets}
For this experiment, we adopt three high-quality datasets that are commonly used in existing research, i.e., SQuAD2, BoolQ, and NarrativeQA. The details of the datasets are shown in Table~\ref{tab:dataset}. 
% The training sets of these datasets were used to train the UnifiedQA pre-trained model, so we choose their test data as the seed for test case generation. The specifics of these datasets are as follows:

\noindent \textbf{\em BoolQ.} BoolQ ~\cite{clark2019boolq} is a boolean QA dataset where each question requires a simple ``yes'' or ``no'' answer. The questions are drawn from aggregated queries made to the Google search engine, and the contextual information is sourced from Wikipedia. 

\noindent \textbf{\em SQuAD2.} SQuAD2 ~\cite{rajpurkar2018know} is a QA dataset where answers are certain text segments extracted from the provided context. The dataset's questions and answers are generated by crowd workers based on paragraphs from Wikipedia. If a question cannot be answered, it is labeled with `\textlangle No Answer \textrangle'.

\noindent \textbf{\em NarrativeQA.} NarrativeQA~\cite{kovcisky2018narrativeqa} is an abstractive QA dataset where answers are given in free-form text, extending beyond concise spans from the context. This dataset, derived from books and movie scripts, is designed to test reading understanding, especially with long and intricate reference texts.
\subsubsection{Experiment Configurations}
We utilized a server equipped with dual Intel(R) Xeon(R) Platinum 8260 CPUs, eight GeForce RTX 2080 Ti GPUs, 502.6 GB of memory, and a 5 TB RAID1 disk array. The large language model, ChatGPT-4o-mini, was accessed through an API interface.
\begin{table*}[t]
\belowrulesep=0pt
\aboverulesep=0pt
    \centering
    \caption{Reported bugs and true positive, $\approx{}$ represents the estimated number of true bugs, calculated by multiplying the total number of detected bugs by the true positive rate.}
    \label{tab:bugsandTP}
    % \begin{adjustbox}{width=0.5\textwidth}
    \begin{tabular}{l|c|c|c|c|c|c|c|c|c|c|c|c} % 设置每列的宽度并允许自动换行
    \toprule
        \multirow{2}{*}{Dataset} & \multicolumn{3}{c|}{\ourtool-Entity} & \multicolumn{3}{c|}{\ourtool-Relation} & \multicolumn{3}{c|}{QAQA} & \multicolumn{3}{c}{QAAskeR-plus} \\ \cline{2-13}
                                & \# Bugs & TP & \# True Bugs & \# Bugs & TP & \# True Bugs & \# Bugs & TP & \# True Bugs& \# Bugs & TP & \# True Bugs\\ \midrule
        BoolQ                  & 17,414 & \textbf{94\%}  &$\approx{\textbf{16,369}}$          & 6,699 & 86\% &$\approx{5,761}$          & 2,423 & 62\%  &$\approx{1,502}$ & 1,760 & 34\%  &$\approx{598}$        \\ 
        SQuAD2                 & 4,878  & \textbf{81\%}  &$\approx{\textbf{3,951}}$        & 2,589 & 74\% &$\approx{1,916}$         & 12,829 & 23\%  &$\approx{2,951}$ & 4,390 & 73\%  &$\approx{3205}$      \\ 
        NarrativeQA            & 2,512  & \textbf{82\%}  &$\approx{\textbf{2,060}}$        & 2,827 & 82\%  &$\approx{2,318}$        & 1,985  & 29\%   &$\approx{576}$ & 2,746  & 74\%   &$\approx{2032}$     \\ \bottomrule
    \end{tabular}
    % \end{adjustbox}
\end{table*}

\begin{table}[tb]
    \centering
    \caption{Proportion of Hallucinated Questions (Content Violation and Instructional Violation)}
    \label{tab:Hallucinated}
    \resizebox{\columnwidth}{!}{  % 自动缩放宽度适应页面
    \begin{tabular}{l|cc|cc|cc|cc}
        \toprule
        \multirow{3}{*}{Dataset}
        & \multicolumn{4}{c|}{\ourtool-Entity} 
        & \multicolumn{4}{c}{\ourtool-Relation} \\
        \cline{2-9}
        & \multicolumn{2}{c|}{Raw} & \multicolumn{2}{c|}{Filtered} 
        & \multicolumn{2}{c|}{Raw} & \multicolumn{2}{c}{Filtered} \\
        \cline{2-9}
        & CV & IV & CV & IV & CV & IV & CV & IV \\
        \midrule
        BoolQ       & 0.14 & 0.04 & 0.02 & 0.01 & 0.13 & 0.47 & 0.03 & 0.02 \\
        SQuAD2      & 0.21 & 0.01 & 0.03 & 0.00 & 0.18 & 0.55 & 0.02 & 0.00 \\
        NarrativeQA & 0.10 & 0.03 & 0.02 & 0.00 & 0.26 & 0.42 & 0.03 & 0.01 \\
        \bottomrule
    \end{tabular}
    }
\end{table}

\begin{table*}[t]
\belowrulesep=0pt
\aboverulesep=0pt
    \centering
    \caption{Quality of the generated questions}
    \label{tab:naturalness}
        % \begin{adjustbox}{width=\textwidth}
 % 按页面宽度缩放表格
    \begin{tabular}{l|c|c|c|c|c|c|c|c|c|c|c|c|c|c|c|c} % 设置每列的宽度并允许自动换行
        \toprule
        \multirow{2}{*}{Dataset} & \multicolumn{4}{c|}{\ourtool-Entity} & \multicolumn{4}{c|}{\ourtool-Relation} & \multicolumn{4}{c|}{QAQA} & \multicolumn{4}{c}{QAAskeR-plus}\\ \cline{2-17}
         & CR1 & CR2  & CR3 & Avg  
         & CR1 & CR2  & CR3 & Avg
         & CR1 & CR2  & CR3 & Avg
         & CR1 & CR2  & CR3 & Avg\\ \midrule
        BoolQ                  & 4.18 & 4.37  & \textbf{4.80} & 4.45 & \textbf{4.44} & \textbf{4.61} & 4.71 & 4.59 
        & 2.60 & 1.93 & 3.27  & 2.60  & 2.33 & 2.97 & 3.44  & 2.91       \\ 
        SQuAD2                 & 4.17 & 4.29  & \textbf{4.80} & 4.42 & \textbf{4.50} & \textbf{4.78} & 4.71 & 4.66  
        & 2.43 & 1.89 & 2.99  & 2.24  & 3.74 & 4.27 & 4.69  & 4.23      \\ 
        NarrativeQA            & 3.69 & 3.84  & \textbf{4.73} & 4.08 & \textbf{3.94} & \textbf{4.59} & 4.61 & 4.38
        & 2.55 & 1.93 & 3.47  & 2.65  & 3.1 & 3.76 & 4.17  & 3.68     \\ \bottomrule
    \end{tabular}
        % \end{adjustbox}
\end{table*}
\subsection{Capability of bug detection}
\label{sec:bugdetection}
\noindent \textbf{RQ1: How is the bug detection capability of \ourtool{}?} 
We first evaluate the effectiveness of \ourtool{} in detecting bugs. We used  \ourtool{}, QAQA and QAAskeR-plus methods to generate test cases from the same seed datasets and then sent the generated test cases to the software under test, UnifiedQA. We then counted the number of bugs detected by each method. The specific numbers are shown in Table~\ref{tab:bugsandTP}. Notably, we explicitly report the results of bug detection with questions generated with entity information (\ourtool{}-Entity) and questions generated with relation information (\ourtool{}-Relation) by \ourtool{}.
% was divided into two parts for display based on the types of information extracted, i.e., entity information and relation information.

From the results, it can be seen that \ourtool{} detected significantly more defects than QAAskeR-plus on all three datesets and more defects than QAQA on the BoolQ and NarrativeQA datasets. However, on the SQuAD2 dataset, our method detected fewer defects than QAQA. This is due to the reason that the SQuAD2 dataset contains a large number of questions with the result of `\textlangle No Answer \textrangle', which have few corresponding context information. 
% due to the large number of unanswerable questions in SQuAD2, which contain less contextual information. 
Since QAQA  generates test cases primarily based on the questions, while our method extracts entities/relations from the context as the ground truth for question generation, lacking of context in the SQuAD2 dataset led to a lower number of bugs detected by \ourtool{} compared to QAQA. However, only 23\% of the bugs identified by \qaqa{} on SQuAD2 are real bugs, whereas \ourtool{} achieves a significantly higher ratio of 81\%. Therefore, \ourtool{} identifies substantially more real bugs compared to \qaqa{} on \squad{}. 
% , that is 3,951 (entity-based) vs 2,951.

We evaluated the authenticity of detected bugs, as reflected by the true positive rate. Since all methods identified a large number of defects, it was impractical to manually label every instance. Consequently, following the approach of QAQA, we randomly selected and annotated 100 defects for each method on each dataset, resulting in a total of 1,200 labeled instances.

Each instance was independently labeled by two experienced researchers with strong English proficiency. 
% A consensus is achieved if two annotators provide the same label to the same annotation  task, directly or through discussions.  
A detected bug was labeled as a true positive if it satisfied the following criteria: (1) the question is coherent, natural, contextually relevant, and clearly understandable to the annotator; (2) the ground truth answer to the question can be found within the context; and (3) the answer provided by the system under evaluation is semantically inconsistent with the ground truth answer. If any of these criteria were not met, the instance was labeled as a false positive. 
It is important to note that, in this paper, the term “false positive” encompasses not only cases where the model does not actually make an error, but also situations where the test cases themselves are flawed. For example, this includes cases in which the generated questions are unnatural or irrelevant to the provided context. This broader definition is adopted because the goal of test case generation is to simulate real user queries. Questions that are unnatural or disconnected from the context are unlikely to appear in practical applications and, therefore, do not contribute meaningful evaluations of model performance. Defects identified by such invalid test cases lack practical value for improvement and are considered false positives.  
% invalid. As a result, this paper classifies defects arising from such invalid test cases as false positives and includes them in the calculation of defect report accuracy.

% positive if then answer provided by the software under test is semantically different from the golden answer reflected by the context information. Otherwise, a true positive label is provided.  
% The annotators labeled defects meeting the above two criteria as True Positives, otherwise as False Positives. 
Cohen’s kappa coefficient between the two sets of labeled data was 0.75, indicating substantial agreement between annotators. All disagreements were resolved through in-depth discussion to achieve a final consensus. We calculated the proportion of True Positives for each set and recorded the results in Table~\ref{tab:bugsandTP}. Furthermore, we estimate the number of true detected bugs by multiplying the total detected bugs by the true positive rate.
As shown in Table~\ref{tab:bugsandTP}, the proportion of true positives among the detected bugs is significantly higher for \ourtool{} compared to \qaqa{} (83\% vs 38\% on average) and to QAAskeR-plus (83\% vs 60\% on average). 
% Based on the true positive rate, we can estimate the number of true bugs by multiplying the number of total bugs by the true positive rate. 
Therefore, the number of true bugs identified by \ourtool{} surpasses that of \qaqa{} and QAAskeR-plus, indicating that the bugs detected by our approach are more credible and more likely to be real bugs. These results demonstrate that \ourtool{} outperforms these two methods in bug detection, achieving a higher number of detected bugs and a superior true positive rate.
% {\bfseries Answer to RQ1}\\
% In scenarios with sufficient contextual information, \ourtool{} is capable of detecting a large number of bugs with a high degree of authenticity, outperforming the baseline method, QAQA
% \subsection{Comparison with state-of-the-art approaches}

\subsection{Effectiveness of Quality Assurance Approach}
\noindent \textbf{RQ2: How effective is the approach for ensuring the quality of \llm{}'s generated outputs?} 
To enhance reliability and ensure the quality of outputs, we propose using consistency verification and constraint checking to filter out problematic tests. This study investigates the effectiveness of these filtering methods and examines whether the remaining test cases meet the specified requirements.
We assess the impact of our filtering methods on improving the quality of results generated by LLMs. Specifically, we compare the data before and after applying the filtering techniques to evaluate their effectiveness.

% Since the number of generated questions is so large that can not all be labeled, 
We randomly selected 100 questions from each type of data for manual labeling, resulting in a total of 1,200 labeled instances. We follow the same labeling process and criteria as that of section~\ref{sec:bugdetection}. 
% Similar to the labeling method in Research Question 1, we duplicated the data into two sets and labeled them independently by two researchers. The labeling criteria are as follows:\\
% 1)Whether the question is clearly stated and easily understandable.\\
% 2)Whether the standard answer is the correct answer for the generated question.\\
% Questions that meet both criteria were labeled as correct, while those that did not were labeled as incorrect. 
The Cohen's kappa coefficient for the two sets of labeled data was 0.76, and all disagreements were resolved through further in-depth discussion to achieve a final consensus. We calculated the proportion of correct questions for each set and recorded the data in Table~\ref{tab:proportion}.

It can be observed that generating questions based on relational information is more challenging for LLMs, as it involves querying the relationship part of the triples. As a result, the proportion of correct questions for relational information without filtering is low (33.0\% on average). On the other hand, generating questions based on entity information is relatively easier, leading to a higher proportion of correct questions (82.3\% on average). Overall, after applying our designed filtering methods, the precision of the generated questions improved significantly (97.3\% on average for entity-based questions, and 96.3\% on average for relational questions). 
\begin{table}[tb]
    \centering
    \caption{Proportion of correct questions}
    \label{tab:proportion}
    % \resizebox{0.85\linewidth}{!}{
    \begin{tabular}{l|c|c|c|c}\toprule % 设置每列的宽度并允许自动换行
        \multirow{2}{*}{Dataset} & \multicolumn{2}{|c|}{\ourtool-Entity} & \multicolumn{2}{|c}{\ourtool-Relation} \\ \cline{2-5}
                                & Raw & Filtered  & Raw & Filtered   \\ \midrule
        BoolQ                  & 0.82 & 0.97           & 0.40 & 0.95          \\ 
        SQuAD2                 & 0.78  & 0.97          & 0.27 & 0.98          \\ 
        NarrativeQA            & 0.87  & 0.98          & 0.32 & 0.96           \\ 
    \bottomrule
    \end{tabular}
    % }
\end{table}
% {\bfseries Answer to RQ2}\\
The filtering method we proposed effectively improve the precision of generated questions, improving the quality of the test cases.

Based on our analysis of the erroneous questions, we find that hallucinations generated by large language models during the question generation process are the primary cause of these errors. Since the task assigned to the model is to generate questions based on the answer and a specific context, the hallucinations observed in this process is mainly faithfulness violations—that is, the generated questions fail to properly follow the instructions provided in the prompt. In our experimental results, these hallucination issues can be further categorized into two distinct types of faithfulness violations:

\begin{itemize}
\item \textbf{Content Faithfulness Violation}: This type refers to inconsistencies between the generated question and the provided background content. It encompasses cases where the question introduces facts, entities, or relationships that do not appear in the context, or misrepresents the original content—for example, by introducing incorrect causal relationships, contradictions, or relying on external knowledge not present in the context.
\item \textbf{Instructional Faithfulness Violation}: This type refers to violations of the constraints specified in the prompt. Common cases include answer leakage (i.e., where the answer is directly stated or implied in the question), asking about information beyond the provided context, or failing to adhere to the required question format—such as generating incomplete questions or those with an ambiguous focus.
\end{itemize}

Based on these criteria, we conducted a further experimental analysis to classify the errors caused by fidelity violations. The results of this classification are presented in Table~\ref{tab:Hallucinated}.

As illustrated in Table~\ref{tab:Hallucinated}, faithfulness violations in question generation show two findings. Firstly, instructional faithfulness violations (IV) occur significantly more frequently in relation-type questions than in entity-type questions, particularly in the Raw setting. For example, in the SQuAD2 dataset, the IV rate for relation questions reaches 0.55, compared to only 0.01 for entity questions. This suggests that relation-based question generation poses greater challenges for large language models, as it demands strict compliance with prompt-level constraints, such as preventing answer leakage and ensuring question completeness.

Second, the application of filtering substantially reduces both content (CV) and instructional (IV) violation rates across all datasets and question types. This finding demonstrates that many faithfulness-related errors originate from test cases that are poorly constructed or misaligned with the context, and that adopting a quality assurance approach is effective for improving test quality and reducing invalid questions.
\subsection{Quality of the Generated Questions}
\noindent \textbf{RQ3: What is the quality of the questions we generate?}
% 
% This experiment evaluates the quality of the questions generated by \ourtool{} and QAQA. 
% As the most typical usage scenario of a QA software is that users raise questions based on the given context information, the questions we generated should be able to simulate this usage scenario to achieve good testing effectiveness. 
% We assess the quality through a scoring system based on three evaluation criteria, as outlined below: (1) Whether the question is phrased in a way that a human would ask; (2) Whether the question has a focused topic. More that one topic occurring in the same question makes the it ambiguous and hard to understand; (3) The answer and primary topic in the question should all appear in the context.
This experiment evaluates the quality of questions generated by \ourtool{} and QAQA. Since the primary usage scenario of a QA system involves users posing questions based on the given context, the generated questions should realistically mimic this behavior to enable effective testing. We assess question quality from two complementary perspectives:

\begin{table}[tb]
    \caption{Proportion of Textual Coverage}
    \label{tab:coverage}
    \centering
    % \begin{adjustbox}{width=\columnwidth}
    \begin{tabular}{lccc}\toprule
\textbf{Dataset}  & \textbf{CQ2A} & \textbf{QAQA} & \textbf{QAAskeR-plus} \\ \midrule
BoolQ                   & \textbf{79\%}          & 43\%     & 39\%                     \\ 
SQuAD2                 & \textbf{85\%}          & 29\%    & 53\%                      \\ 
NarrativeQA           & \textbf{51\%}          & 6\%      & 49\%            \\ \bottomrule

\end{tabular}
% \end{adjustbox}
\end{table}
\vspace{1mm}
\noindent\textbf{\em (1) Naturalness Evaluation}:
We adopt a human-centric scoring system based on the following three criteria:

\begin{itemize}
\item Natural Phrasing (CR1): Whether the question is phrased in a way that a human would naturally ask.
\item Topical Clarity (CR2): Whether the question has a clear and focused topic—questions that involve multiple unrelated topics are often ambiguous and difficult to interpret.
\item Contextual Grounding (CR3): Whether both the answer and the main subject of the question appear within the provided context.
\end{itemize}

For each criterion, we designed a 0–5 scoring scale with specific guidelines for each score. GPT-4o is utilized as the evaluation tool. The scoring rules and standards are embedded in a prompt, which guides the model to assess each generated question in the context of its source passage. The model assigns a score for each criterion and provides an explanation for its evaluation, after which an overall average score is computed across the three criteria.
We randomly sampled 100 contexts from each dataset and evaluated the naturalness of the generated questions according to the above criteria. We then calculated the average score for each criterion as well as the overall mean score. The final results are shown in Table~\ref{tab:naturalness}.

As shown in the table, \ourtool{} consistently outperforms both QAQA and QAAskeR-plus across all datasets and criteria. Specifically, the questions generated by \ourtool{} (both \ourtool{}-Entity and \ourtool{}-Relation) achieve significantly higher average scores. For example, on the BoolQ dataset, the average scores are 4.61 and 4.44 for \ourtool{}-Relation and \ourtool{}-Entity, respectively, compared to only 3.27 for QAQA and 2.91 for QAAskeR-plus. Similar trends are observed for SQuAD2 and NarrativeQA.
These results indicate that the questions generated by \ourtool{} are more natural, focused, and closely tied to the context. In other words, they better simulate authentic user behavior, enabling more realistic and effective evaluation of QA systems.

\vspace{1mm}
\noindent\textbf{\em (2) Coverage Evaluation}:
To further assess the comprehensiveness of generated questions, we evaluate the extent to which the source context is covered. Greater coverage reflects a more thorough examination of the context. For this purpose, \ourtool{} employs GPT-4o to identify which segments of the context are addressed by the generated questions. Specifically, all questions and the complete context are input into a prompt, instructing the model to match each question to its corresponding segment in the context. Coverage is then computed as the ratio of the total length of all matched segments to the overall length of the context.

According to the results, CQ2A achieves substantially higher textual coverage than QAQA and QAAskeR-plus across all datasets. For instance, on the SQuAD2 dataset, CQ2A attains 85\% coverage, while QAQA and QAAskeR-plus achieve only 29\% and 53\%, respectively. Similarly, on the BoolQ dataset, CQ2A achieves 79\% coverage, clearly outperforming QAQA (43\%) and QAAskeR-plus (39\%). Even on the more challenging NarrativeQA dataset, CQ2A still reaches 51\% coverage, compared to just 6\% for QAQA and 49\% for QAAskeR-plus.
These findings demonstrate that the questions generated by \ourtool{} cover a much broader portion of the source context, indicating that its test cases more comprehensively and realistically reflect real-world user behavior. \ourtool{} produces higher-quality test cases than both QAQA and QAAskeR-plus, contributing to more effective QA system evaluation.

% Together, these two evaluations provide a comprehensive view of question quality, considering both linguistic naturalness and contextual completeness.

% \begin{itemize}
%     \item Whether the question is phrased in a way that a human would ask.
%     \item Whether the question has a focused topic. More that one topic occurring in the same question makes the it ambiguous and hard to understand. 
%     \item The answer and primary topic in the question should all appear in the context. 
% \end{itemize}
% Focus on the main semantic points of the question, ignoring contextual or supplementary information. Please summarize how many core semantic points the question has. A good question should have only one main semantic point.

% which indicates that the test cases have higher value in the testing process.

\subsection{Effectiveness for Improving the QA Software}
\noindent \textbf{RQ4: Can the test cases generated by \ourtool{} help improve the QA software? }
Lastly, we evaluate to what extent our method can improve the performance of question-answering software. We fine-tuned the UnifiedQA model (small) using the generated test cases as the training set and evaluated its error rate on the same test set before and after fine-tuning. Specifically, for each dataset, we randomly selected 10,000 test cases generated with relation and entity information as the training set, 1,000 cases as the validation set, and 1,000 cases as the test set, and make sure that any duplicate cases were removed. We finetuned the UnifiedQA model with the training set and saved the fine-tuned model after 1,000 epochs.

The data of the bug occurrence rate before and after fine-tuning the model is shown in Table~\ref{tab:finetune}. As we can see, compared to the original model, the fine-tuned model significantly reduced the error rate across all three datasets. The improvement is most noticeable on the NarrativeQA dataset, where the error rate decreased by 33.2\%.
\begin{table}[t]
    \caption{Error rate of original model and finetuned model}
    \label{tab:finetune}
    \centering
    % \begin{adjustbox}{width=\columnwidth}
    \begin{tabular}{lcc}\toprule
\textbf{Dataset}  & \textbf{Original model} & \textbf{Finetune model} \\ \midrule
BoolQ                      & 42.8\%                      & 16.8\%                 \\ 
SQuAD2                 & 34.4\%                    & 12.5\%                \\ 
NarrativeQA           & 50.7\%               & 17.5\%               \\ \bottomrule
\end{tabular}
% \end{adjustbox}
\end{table}
\section{Related Work}
\label{sec:relatedwork}
% \subsection{QA Software Testing}
Existing research in QA testing can be divided into \textit{reference-based techniques} and \textit{automated testing techniques}. 

\noindent \textbf{\em Reference-based techniques} rely on manual annotation, and many standard datasets have been created, including BoolQ~\cite{clark2019boolq}, NatQA~\cite{kovcisky2018narrativeqa}, SQuAD2~\cite{rajpurkar2018know}, MultiSpanQA~\cite{li2022multispanqa}, Quoref~\cite{dasigi2019quoref}, etc.
Reference-based techniques demand significant human effort, making automated techniques the primary focus of researchers. 

\noindent \textbf{\em Automated testing techniques} can be divided into two types based on the data sources for generation. \textit{The first type involves generating test cases based on information from external knowledge bases.} \cite{wang2024knowledge} extracted entity relationships from established large knowledge graphs, designed reasoning logic to merge two sets of entity relationships, and generated Yes/No questions based on these relationships. \textit{The other type involves mutation testing based on existing QA datasets.} \cite{chen2021testing}, \cite{xie2023qaasker+} proposed an automated QA software testing technique called QAAskeR-plus, which synthesizes declarative facts based on the QA results of the software being tested. These facts are then syntactically analyzed, and potential answers are selected. Subsequently, the UniLm~\cite{dong2019unified} model is used to regenerate new questions based on the facts and answers, and these new questions and answers are combined as new test samples. \cite{liu2022qatest} designed a complete fuzz testing framework for QA software, with a multi-level semantic mutation method ranging from words to sentences. To ensure the diversity and effectiveness of the generated samples, the author evaluates the quality of the generated test cases in terms of coverage and ambiguity, selecting those that meet the criteria. \cite{shen2022natural} proposed a series of sentence-level mutation relations, arguing that when irrelevant sentences are added to the question and context, their corresponding answers should remain consistent with the original ones. On this basis, five types of semantic-guided mutation methods were developed, generating new test samples based on the original dataset for QA software testing. These works focus on modifying the original dataset’s questions and contexts to create new test cases. In contrast, our approach generates new questions based on information extracted from the context, which has a higher coverage to the context.

% \ls{(1) The paragraph is too long, separate it somewhere. (2) Add discussions on the differences of the related work with \ourtool. }
\section{Conclusion}
\label{sec:conclusion}
In this work, we propose \ourtool—a testing method for QA software that leverages the power of large language models (LLMs) to generate questions based on the ground truth answer and the context from which it is derived. By considering the context, \ourtool{} is able to generate more diverse questions that capture more contextual information. Additionally, LLMs enable the creation of natural, human-like questions, making the test cases generated by \ourtool{} more effective in evaluating QA software. Our extensive evaluation demonstrates that \ourtool{} outperforms state-of-the-art QA testing approaches in terms of bug detection, true positive rate, context coverage and the naturalness of the generated questions.

\bibliographystyle{IEEEtran}
\bibliography{refs}

\end{document}